\documentclass[a4paper, oneside, twocolumn, notitlepage, 10pt]{extarticle_ecoc}
\usepackage{ecoc}
\usepackage{wrapfig}
\usepackage{subcaption}
\usepackage{amsmath,amssymb,amsfonts}
\usepackage{graphicx}

\hypersetup{bookmarksdepth=-2}

\begin{document}
\selectlanguage{english}

\title{Analysis of the Coherent Contributions to Nonlinear Interference Generation within Disaggregated Optical Line Systems}
\author{
    Elliot London\textsuperscript{(1)}, Emanuele Virgillito\textsuperscript{(1)},
    Andrea D'Amico\textsuperscript{(1)}, Antonio Napoli\textsuperscript{(2)},
    Vittorio Curri\textsuperscript{(1)}
}
\maketitle

\begin{strip}
 \begin{author_descr}
 
   \textsuperscript{(1)} Department of Electronics and Telecommunications, Politecnico di Torino, Corso Duca degli Abruzzi 24, 10129 Torino, Italy,
   \textcolor{blue}{\uline{elliot.london@polito.it}}
   
   \textsuperscript{(2)} Infinera, St.-Martin-Strasse 76, 81541 M\"unchen, Germany,
   
 \end{author_descr}
\end{strip}

\setstretch{1.1}

\begin{strip}
  \begin{ecoc_abstract}
    % NOTE: Don't use a blank line here but start abstract right away to avoid an extra line break
    Through a physical layer simulation study we highlight that the coherent accumulation of nonlinear interference becomes non-negligible for optical networks operating within high symbol rate transmission scenarios. This initiates a discussion on how to accurately plan and model optical networks, starting from the physical layer.
  \end{ecoc_abstract}
\end{strip}
%
%%%
\section{Introduction}
The current and future requirements from optical line system (OLS) operators, cloud service providers and end users are producing a progressive yearly increase in demand for higher transmission rates and larger optical network capacities and throughput~\cite{cisco2019,varghese2018next,ferrari2020assessment}. 
%
% \commentAnt{Another reason is that operators wish to reduce the number of cards. 2$\times$400G is supposed to be still more expensive than 800G, and also important is the management of the network: the lower the number of cards, the easier. We should also mention that, see our Infinera JLT on DSP ASIC 2020, that if one move to single wavelength with high symbol rate, we might end up in other problem. So at the end, I believe that the solution will be digital subcarrier but really closely packed.}
%
Correspondingly, optical networks are shifting to a format that enables a greater degree of flexibility with respect to wavelength assignment, transmission standards and symbol rates, representing a shift towards a disaggregated structure~\cite{roberts2017beyond,sambo2015next,sun2020800g}.
A fully disaggregated network approach requires all OLS components to be considered independently~\cite{auge2019open}, as they may be controlled by different network providers and operators. In this scenario the lightpath (LP) history may not be known, presenting a challenge when attempting to model the physical layer: the nonlinear interference (NLI) produced during propagation is coherent~\cite{icton2019,ONDM2020}, meaning that the NLI accumulation depends upon the configurations of all previously crossed fiber spans.
This problem is further compounded by the presence of alien wavelengths, which are signals originating from third-party equipment that must be properly managed during network planning, complicating path assignment and, in turn, the LP history~\cite{proietti2019experimental}.
NLI accumulation is often computed using a fully incoherent model, such as the extended family of Gaussian noise (GN) models~\cite{poggiolini2012gn,egn}, which have proven to be accurate and conservative for current operational symbol rates, at the least for full C-band occupation~\cite{nespola2013gn}.
As symbol rates increase, so too does the coherent contribution to the total NLI accumulation~\cite{ONDM2020}, reducing the accuracy of a fully incoherent model and negatively impacting the ability of OLS controllers to measure quality of transmission (QoT) degradation, set operational margins and adapt to changing network demands.
To ensure optimal prediction of QoT degradation within modern optical networks it is therefore important to determine the magnitude of the coherent contribution to the overall generalized signal-to-noise-ratio (GSNR) degradation, in order to achieve maximum operational capacities, satisfy future end user requirements and preserve a conservative margin.
Within this paper we present the results of an extensive simulation campaign using an internally-developed, MATLAB\textsuperscript{\textregistered}-based software framework~\cite{pilori-ffss}, which is used to assess the ability for an optical network to be fully disaggregated starting from the physical layer. In this framework, we analyze the coherent contribution to the overall QoT degradation. Consequently, we discuss the impact of this contribution in the case of future network planning, in particular for optical networks that utilize high symbol rate transmission.
%
%%%
\section{NLI Generation in a Disaggregated OLS}
The QoT degradation in optical networks that utilize coherent technology is usually quantified using the GSNR~\cite{filer18multi}. The GSNR takes into account the contributions of both the amplified spontaneous emission (ASE) produced by optical amplifiers and the NLI that is produced due to nonlinear propagation through the optical fiber.
Considering a cascade of $N$ optical domains, the total GSNR at the termination of the final domain is given by
\begin{equation}
    \mathrm{GSNR} = \left(\sum_{i=1}^{N}\mathrm{GSNR}_i^{-1}\right)^{-1}
    \;,
    \label{eq_GSNR}
\end{equation}
where $i$ is the index of the fiber span and $\mathrm{GSNR}_i$ is the GSNR of each optical domain.
The NLI generation depends upon the transmission standards, symbol rates, spacing and insertion powers of all active channels. The ASE noise may be evaluated by characterizing the optical amplifiers within the OLS and is fully separable from the NLI; consequently we focus only upon the generation and accumulation of NLI within this work.
The NLI may be separated into its two constituent contributions: cross-phase modulation (XPM) and self-phase modulation (SPM).
Due to walk-off, the XPM may be can be considered local and incoherent~\cite{ucl14,icton2019} for a wide range of use cases. In contrast, SPM is a coherent and non-local effect, being the primary hindrance to a fully disaggregated physical layer model. Fortunately, it is possible to quantify the maximum SPM value after a given number of spans~\cite{ONDM2020}, providing an upper limit to the SPM accumulation.

It has previously been shown that the SPM and XPM are separable~\cite{dar2013,egn,icton2019} for all realistic use cases~\cite{ucl14,icton2019}, meaning that the total amount of NLI power, $P_{\mathrm{NLI}}$, generated for a single channel under test (CuT) may be calculated as a sum of the individual channel contributions
\begin{equation}
    P_{\mathrm{NLI}} = P_{\mathrm{NLI},0} + \sum_{k \neq 0} P_{\mathrm{NLI},k}
    \;,
    \label{Eq:totNLI}
\end{equation}
where $k$ is the channel index and $P_{\mathrm{NLI},k}$ is the total amount of NLI power for the $k$th channel. In Eq.~\ref{Eq:totNLI}, the first and second terms enclose the entirety of the SPM and XPM effects, respectively.
Assuming spatial separability allows the total NLI power to also be summed on a span-by-span basis, yielding the total nonlinear SNR power, $\textrm{SNR}_{\textrm{NL}}$
\begin{equation}
    \textrm{SNR}_{\textrm{NL}} = \left(\sum_{i=1,k}^{N}
    \textrm{SNR}_{\textrm{NL},i,k}
    \right)^{-1}\;,
    \label{Eq:SNR_NL}
\end{equation}
where $\textrm{SNR}_{\textrm{NL},i,k}$ are the $i$th span and $k$th channel contributions to the total $\textrm{SNR}_{\textrm{NL}}$. An accurate recovery of this total via a superposition of all individual $\textrm{SNR}_{\textrm{NL},i,k}$ contributions enables the total SPM contribution to be analyzed on a pump-by-pump and span-by-span basis.
%
%%%
\section{Simulation Campaign}
\begin{figure}[t]
    \centering
    \includegraphics[width=1\linewidth]{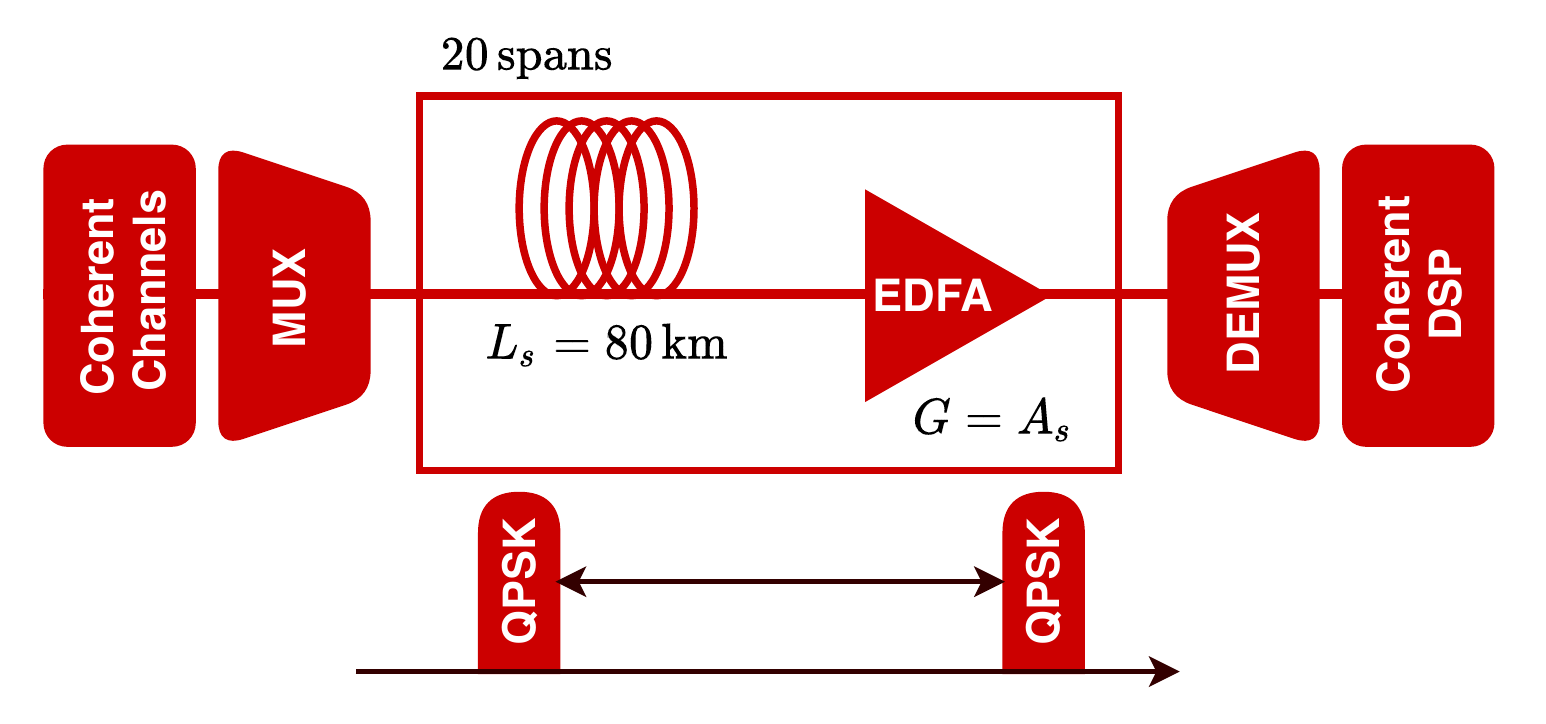}
    % \includegraphics[width=0.225\linewidth]{figures/scattering_diagram.pdf}
    % Scattering diagram not included as space is restricted
    \caption{A sketch of the simulation set-up.}
    \label{fig:OLS_diagram}
\end{figure}

To evaluate the contributions to the total NLI in a wide range of realistic use cases we conducted a simulation campaign over a wide set of OLS configurations and signal parameters, comparing, for each, two distinct simulation scenarios; which we label as the \emph{full-spectrum} and \emph{superimposed} cases, with a sketch of the overall simulation flow shown in Fig.~\ref{fig:OLS_diagram}.
The full-spectrum scenario consists of the propagation of a multi-channel, wavelength-division multiplexed (WDM) signal, representing the target system under investigation.
The superimposed case performs multiple separate pump-and-probe (P\&P) simulations that, in turn, consider the propagation of the channel under test (the probe) and another channel within the WDM grid (the pump).
By creating a superposition of each of these individual P\&P simulations, the result may be compared to the full-spectrum using Eq.~\ref{Eq:totNLI}, allowing the individual NLI and SPM contributions to be investigated.
%
%%%
\section{Results \& Analysis}
\begin{figure}[t]
    \captionsetup{singlelinecheck=false, format=hang, justification=raggedright}
    \centering
    \begin{subfigure}[t]{1\linewidth}
        \includegraphics[width=1\linewidth]{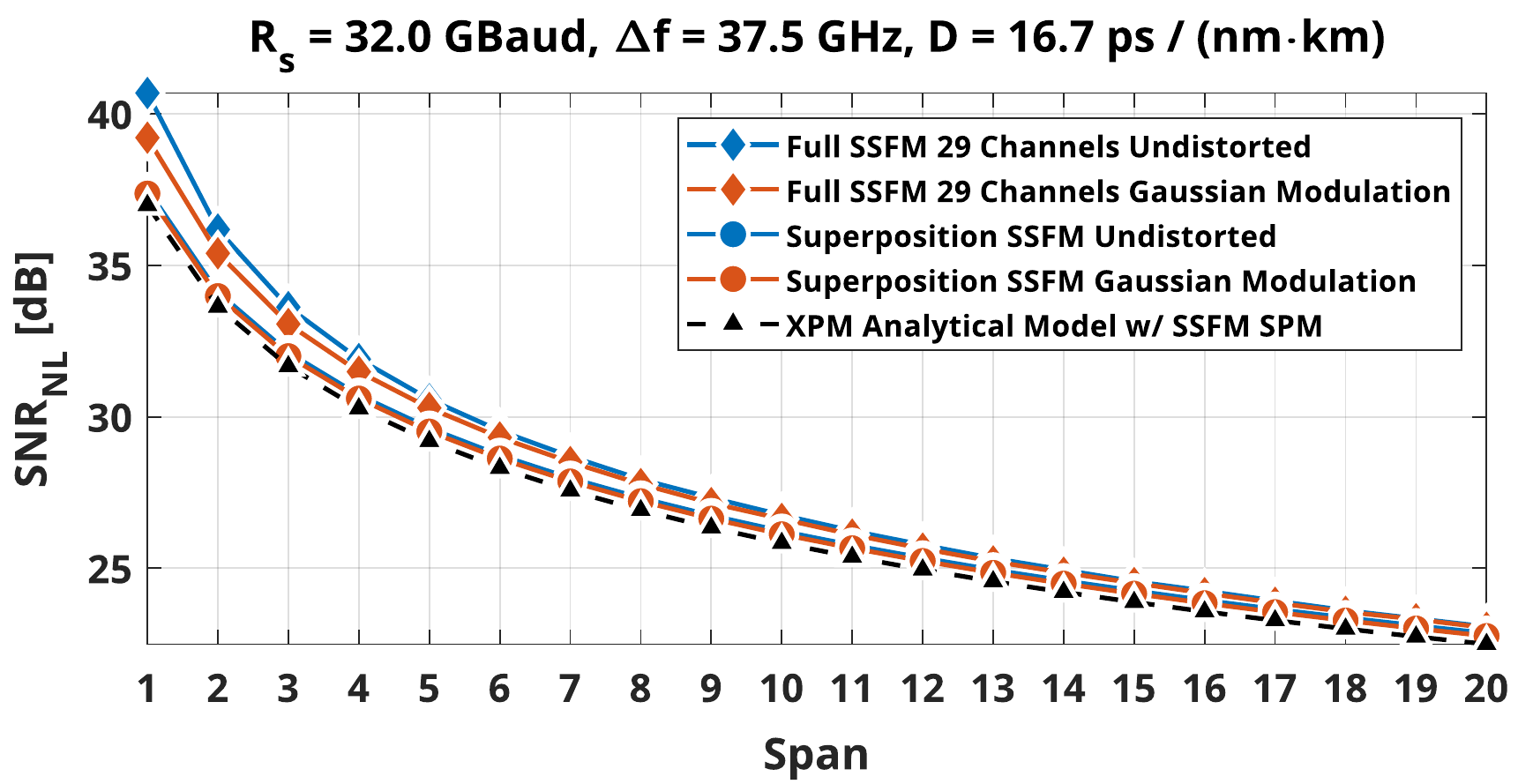}
        \vspace{-45mm}
        \caption{}
        \label{fig:accumulation_a}
    \end{subfigure}
    \begin{subfigure}[t]{1\linewidth}
        \includegraphics[width=1\linewidth]{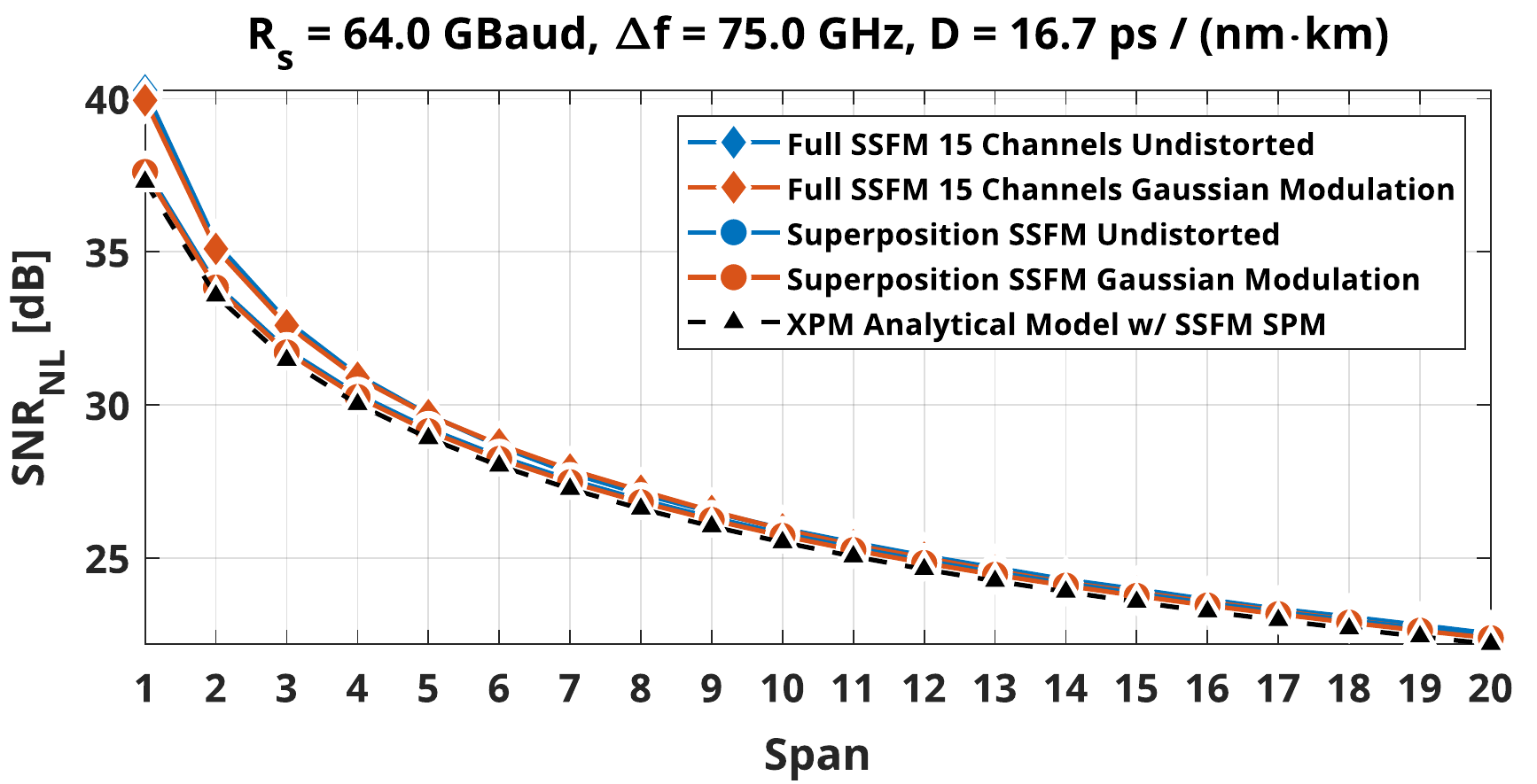}
        \vspace{-45mm}
        \caption{}
        \label{fig:accumulation_b}
    \end{subfigure}
    \caption{The $\mathrm{SNR}_{\mathrm{NL}}$ accumulations for: \textbf{(a)} a widespread use case, and \textbf{(b)} a realistic high symbol rate transmission scenario. A full-spectrum simulation is compared to a superposition of all constituent pump-and-probe simulations, for both predistortion cases.}
    \label{fig:accumulations}
\end{figure}

\begin{figure}[t]
    \captionsetup{singlelinecheck=false, format=hang, justification=raggedright}
    \centering
    \begin{subfigure}[t]{1\linewidth}
        \includegraphics[width=1\linewidth]{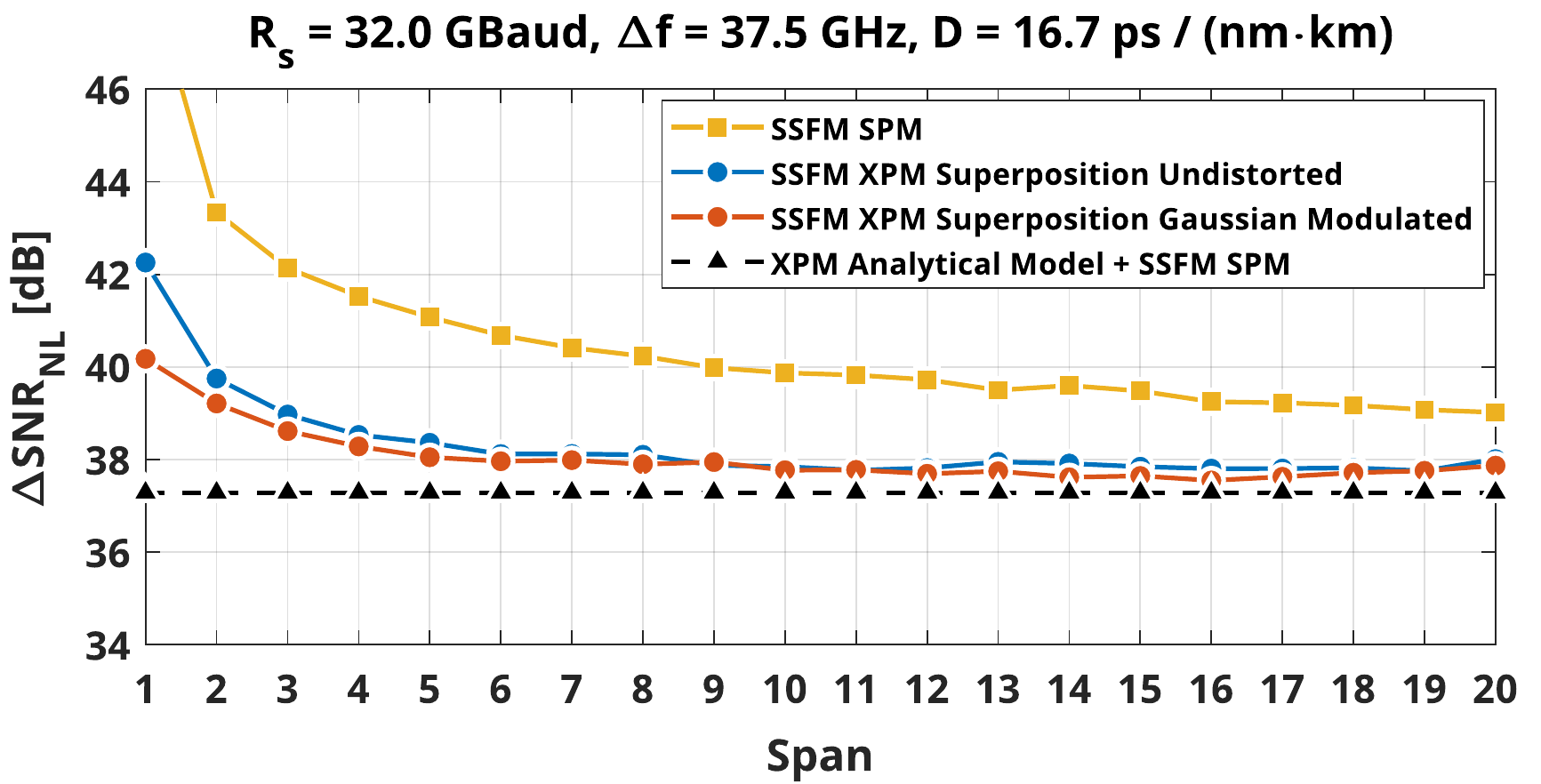}
        \vspace{-45mm}
        \caption{}
        \label{fig:gradient_a}
    \end{subfigure}
    \begin{subfigure}[t]{1\linewidth}
        \includegraphics[width=1\linewidth]{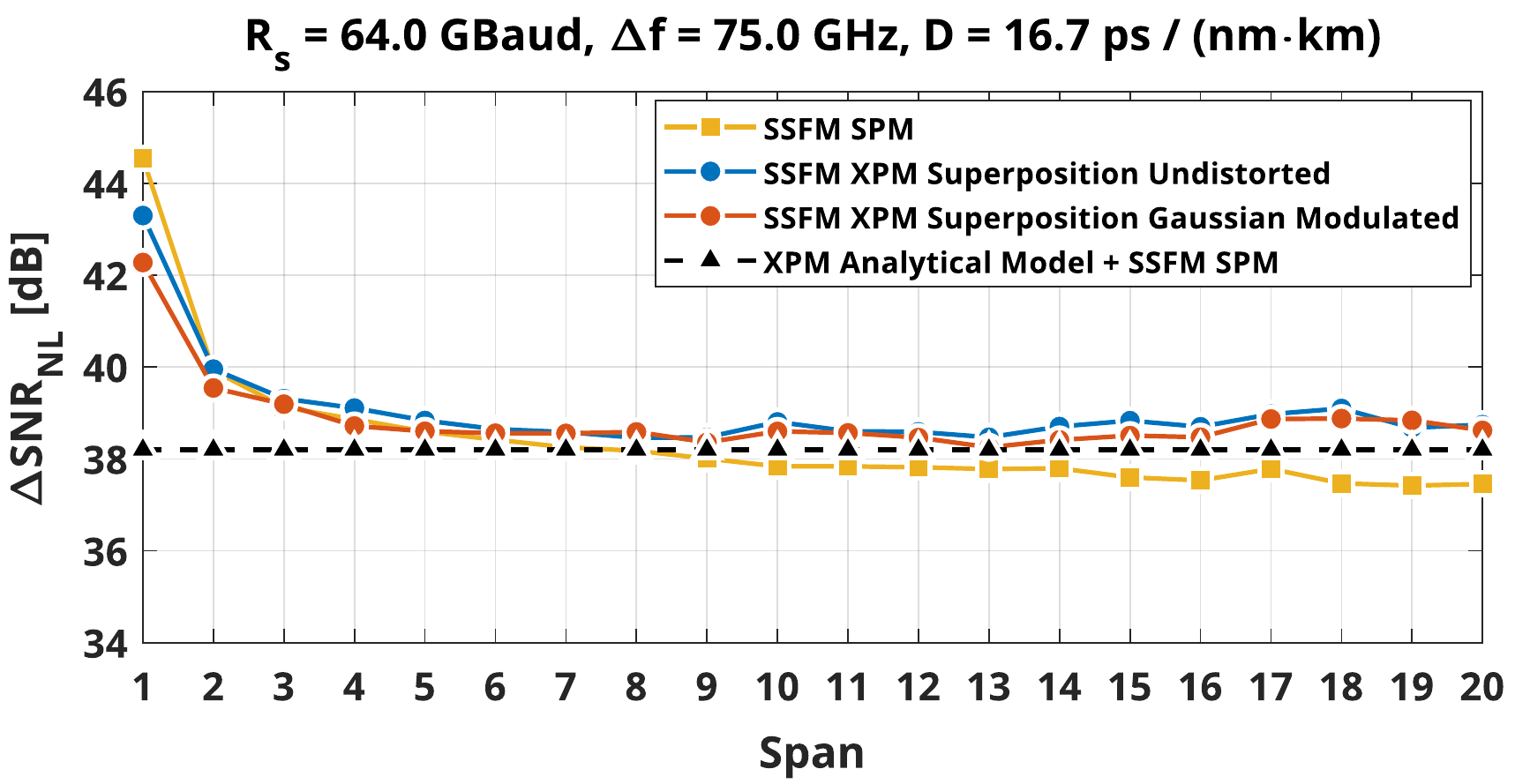}
        \vspace{-45mm}
        \caption{}
        \label{fig:gradient_b}
    \end{subfigure}
    \caption{The $\textrm{SNR}_{\textrm{NL}}$ span-by-span increments, $\Delta\mathrm{SNR}_{\mathrm{NL}}$, for: \textbf{(a)}, a widespread use case and \textbf{(b)}, a realistic high symbol rate transmission scenario. A full-spectrum simulation is compared to a superposition of all constituent pump-and-probe simulations, for both predistortion cases.}
    \label{fig:gradients}
\end{figure}

The simulation campaign encompasses a variety of baud rates, WDM grid spacings and dispersions; within this work we highlight two specific configurations; a widespread use case ($R_s$\,=\,$32.0$\,GBaud, $\Delta f$\,=\,$37.5$\,GHz) and a realistic high symbol rate transmission scenario ($R_s$\,=\,$64.0$\,GBaud, $\Delta f$\,=\,$75.0$\,GHz), both with dispersion values of $D$\,=\,$16.7$\,ps\,/\,(nm$\cdot$km), equal to that of a standard single-mode fiber (SMF).
We launch all scenarios with polarization multiplexed-quadrature phase shift keying (PM-QPSK) modulation applied to the CuT, and for two pump predistortion cases; zero predistortion, representing transmission through a standalone OLS, and Gaussian modulated signals, representing an OLS that is a subsystem of a larger network with significant dispersion having been accumulated along the path of the LP.
Firstly, we present the $\mathrm{SNR}_{\mathrm{NL}}$ accumulations of the full-spectrum simulation and the corresponding superposition in Fig.~\ref{fig:accumulations}, for both predistortion cases. Additionally we include, for reference, the result of an analytical derivation of the XPM contribution to the total NLI, as described in~\cite{icton2019}.
We remark that, for both predistortion cases, the superposition is able to recover the result of the full-spectrum case and serves as an accurate and conservative prediction in all configuration scenarios.
Next, the $\textrm{SNR}_{\textrm{NL}}$ span-by-span increments, $\Delta\mathrm{SNR}_{\mathrm{NL}}$, of these scenarios are presented in Fig.~\ref{fig:gradients}, showing the increments of the superposition, the SPM contribution of the CuT and the aforementioned analytical XPM contribution, for both predistortion cases.
It is visible within these plots that the XPM accumulation contains a transient that quickly vanishes after the first few spans of transmission, before settling to a constantly increasing value, representing fully incoherent accumulation.
Conversely, the SPM continues to accumulate coherently, with the contribution to the overall SNR degradation continuing to increase even after a large number of fiber spans have been crossed.
For the $Rs$\,=\,$32.0$\,Gbaud, $\Delta f$\,=\,$37.5$ and $D$\,=\,$16.7$\,ps\,/\,(nm$\cdot$km) case, the SPM contribution remains small with respect to the total XPM contribution, even at the termination of the final fiber span.
On the other hand, in the $Rs$\,=\,$64.0$\,Gbaud, $\Delta f$\,=\,$75.0$ and $D$\,=\,$16.7$\,ps\,/\,(nm$\cdot$km) case, the SPM contribution becomes dominant after approximately 5 spans of transmission, even accounting for the majority of the NLI contribution at the final fiber span.
Additionally, we highlight that the SPM contribution progressively increases for higher symbol rates, meaning that this issue will be compounded further as transmission rates increase in future generations of coherent systems.
As a result, we stress that, for state-of-the-art optical networks such as those consisting primarily of SMF-type fiber and transmitting signals with a symbol rate of \emph{at least} $Rs=64.0$\,Gbaud, a fully incoherent model is insufficient to accurately model the NLI accumulation.
% Furthermore, this inaccuracy is accentuated when dealing with disaggregated optical networks, on account of the previously mentioned lack of knowledge of the OLS configuration and the history of the LP, increasing the overall required operational error margin.
%
To tackle this issue, it is necessary to consider a model of the total NLI power that properly considers the coherent behaviour of the SPM. We wish to emphasize that a model which considers the total NLI contribution on a channel-by-channel basis satisfies this requirement, permitting an accurate and conservative recovery of the $\textrm{SNR}_{\textrm{NL}}$ and an easier handling of the possibility of alien wavelengths from a network planning point of view.
%
%%%
\section{Conclusions}
Within this work we show that, for coherent optical systems operating in both widespread and high symbol rate transmission scenarios, the total NLI generation for an OLS under full spectral load may be recovered by considering a superposition of all corresponding pump-and-probe contributions. We subsequently show that the contribution of coherent effects to the total NLI becomes dominant for high symbol rate scenarios, such as for $R_s$\,=\,$64.0$\,GBaud and above, highlighting that it is necessity to use a coherent NLI model when calculating the GSNR in optical networks that utilize high symbol rate transmission.
%
%%%
\section{Acknowledgements}
\footnotesize
\begin{wrapfigure}{o}{0.19\linewidth}
    \centering
    \vspace{-3mm}
    \includegraphics[width=1\linewidth]{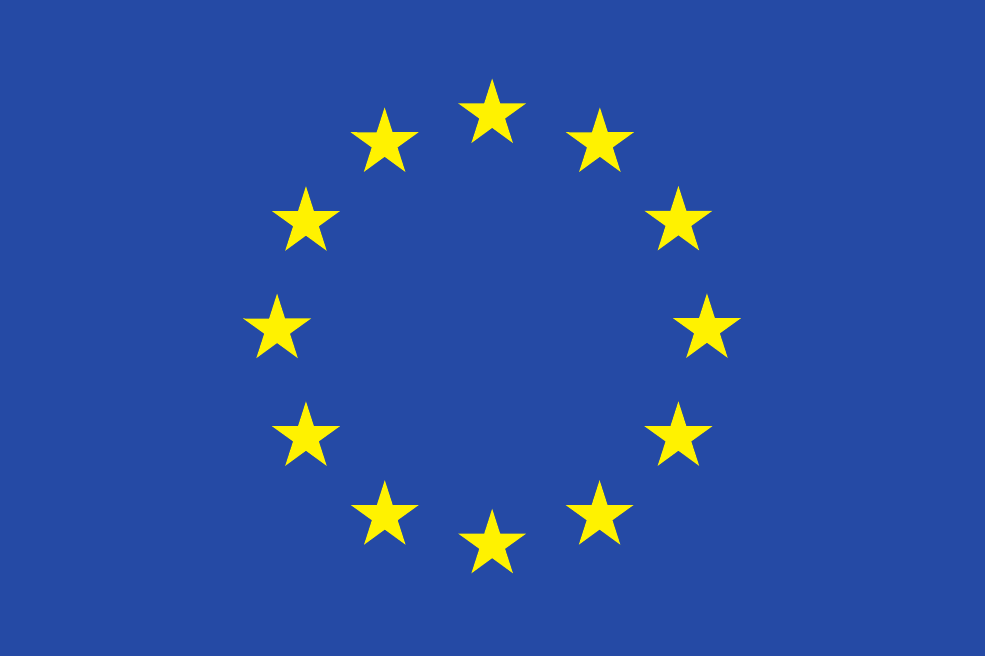}
\end{wrapfigure}
This project has received funding from the European Union's Horizon 2020 research and innovation program under the Marie Sklodowska-Curie grant agreement 814276.
%
%%%
% References may use a fourth page. This additional page is for references only. Footnotes, should appear as footnotes, and should not be included within the references section.
\clearpage
\newpage
\printbibliography

@article{sun2020800g,
  title={800G DSP ASIC Design Using Probabilistic Shaping and Digital Sub-Carrier Multiplexing},
  author={Sun, Han and Torbatian, Mehdi and Karimi, Mehdi and Maher, Robert and Thomson, Sandy and Tehrani, Mohsen and Gao, Yuliang and Kumpera, Ales and Soliman, George and Kakkar, Aditya and others},
  journal={Journal of Lightwave Technology},
  year={2020},
  publisher={IEEE}
}

@article{ferrari2020assessment,
  title={Assessment on the Achievable Throughput of Multi-band ITU-T G. 652. D Fiber Transmission Systems},
  author={Ferrari, Alessio and Napoli, Antonio and Fischer, Johannes Karl and da Costa, Nelson Manuel Simes and D'Amico, Andrea and Pedro, Joao and Forysiak, Wladek and Pincemin, Erwan and Lord, Andrew and Stavdas, Alexandros and others},
  journal={Journal of Lightwave Technology},
  year={2020},
  publisher={IEEE}
}

@misc{cisco2019, 
  author = {Cisco},
  title = {Cisco Visual Networking Index: Forecast and Methodology},
  year = {2018},
  howpublished  = {\url{https://www.cisco.com/c/en/us/solutions/service-provider/visual-networking-index-vni/index.html}},
  note = "[Online; accessed 14-July-2020]",
}

@article{varghese2018next,
  title={Next generation cloud computing: New trends and research directions},
  author={Varghese, Blesson and Buyya, Rajkumar},
  journal={Future Generation Computer Systems},
  volume={79},
  pages={849--861},
  year={2018},
  publisher={Elsevier}
}

@article{proietti2019experimental,
  title={Experimental demonstration of machine-learning-aided QoT estimation in multi-domain elastic optical networks with alien wavelengths},
  author={Proietti, Roberto and Chen, Xiaoliang and Zhang, Kaiqi and Liu, Gengchen and Shamsabardeh, Mohammadsadegh and Castro, Alberto and Velasco, Luis and Zhu, Zuqing and Yoo, SJ Ben},
  journal={Journal of Optical Communications and Networking},
  volume={11},
  number={1},
  pages={A1--A10},
  year={2019},
  publisher={Optical Society of America}
}

@article{nespola2013gn,
  title={GN-model validation over seven fiber types in uncompensated PM-16QAM Nyquist-WDM links},
  author={Nespola, Antonino and Straullu, Stefano and Carena, Andrea and Bosco, Gabriella and Cigliutti, Roberto and Curri, Vittorio and Poggiolini, Pierluigi and Hirano, Masaaki and Yamamoto, Yoshinori and Sasaki, Takashi and others},
  journal={IEEE Photonics Technology Letters},
  volume={26},
  number={2},
  pages={206--209},
  year={2013},
  publisher={IEEE}
}

@article{roberts2017beyond,
  title={Beyond 100 Gb/s: capacity, flexibility, and network optimization},
  author={Roberts, K. and others},
  journal={IEEE/OSA Journal of Optical Communications and Networking},
  volume={9},
  number={4},
  pages={C12--C23},
  year={2017},
  publisher={IEEE}
}

@article{sambo2015next,
  title={Next generation sliceable bandwidth variable transponders},
  author={Sambo, Nicola and others},
  journal={IEEE Communications Magazine},
  volume={53},
  number={2},
  pages={163--171},
  year={2015},
  publisher={IEEE}
}

@ARTICLE{filer18multi,
    author={Filer, Mark and others},
    journal={Journal of Lightwave Technology},
    title={Multi-Vendor Experimental Validation of an Open Source QoT Estimator for Optical Networks},
    year={2018},
    volume={36},
    number={15},
    pages={3073-3082},
    doi={10.1109/JLT.2018.2818406},
    ISSN={1558-2213},
}

@inproceedings{auge2019open,
  title={Open optical network planning demonstration},
  author={Auge, Jean-Luc and Grammel, Gert and Le Rouzic, Esther and Curri, Vittorio and Galimberti, Gabriele and Powell, James},
  booktitle={Optical Fiber Communication Conference},
  pages={M3Z--9},
  year={2019},
  organization={Optical Society of America}
}

@article{dar2013,
  title={Properties of nonlinear noise in long, dispersion-uncompensated fiber links},
  author={Dar, Ronen and Feder, Meir and Mecozzi, Antonio and Shtaif, Mark},
  journal={Optics Express},
  volume={21},
  number={22},
  pages={25685--25699},
  year={2013},
  publisher={Optical Society of America}
}

@article{egn,
  title={EGN model of non-linear fiber propagation},
  author={Carena, Andrea and Bosco, Gabriella and Curri, Vittorio and Jiang, Yanchao and Poggiolini, Pierluigi and Forghieri, Fabrizio},
  journal={Optics express},
  volume={22},
  number={13},
  pages={16335--16362},
  year={2014},
  publisher={Optical Society of America}
}

@article{ucl14,
  title={Adapting transmitter power and modulation format to improve optical network performance utilizing the Gaussian noise model of nonlinear impairments},
  author={Ives, David J and Bayvel, Polina and Savory, Seb J},
  journal={Journal of Lightwave Technology},
  volume={32},
  number={21},
  pages={3485--3494},
  year={2014},
  publisher={IEEE}
}

@inproceedings{pilori-ffss,
  author={D. {Pilori} and M. {Cantono} and A. {Carena} and V. {Curri}},
  booktitle={2017 19th International Conference on Transparent Optical Networks (ICTON)}, 
  title={FFSS: The fast fiber simulator software}, 
  year={2017},
  volume={},
  number={},
  pages={1-4},}

@inproceedings{icton2019,
  title={Observing and Modeling Wideband Generation of Non-Linear Interference},
  author={Virgillito, Emanuele and D'Amico, Andrea and Ferrari, Alessio and Curri, Vittorio},
  booktitle={2019 21st International Conference on Transparent Optical Networks (ICTON)},
  pages={1--4},
  year={2019},
  organization={IEEE}
}

@INPROCEEDINGS{ONDM2020,
  author={A. {D'Amico} and E. {London} and E. {Virgillito} and A. {Napoli} and V. {Curri}},
  booktitle={2020 International Conference on Optical Network Design and Modeling (ONDM)}, 
  title={Quality of Transmission Estimation for Planning of Disaggregated Optical Networks}, 
  year={2020},
  volume={},
  number={},
  pages={1-3},
}

@article{poggiolini2012gn,
  title={The GN model of non-linear propagation in uncompensated coherent optical systems},
  author={Poggiolini, Pierluigi},
  journal={Journal of Lightwave Technology},
  volume={30},
  number={24},
  pages={3857--3879},
  year={2012},
  publisher={IEEE}
}
\end{document}